\documentclass[twocolumn,showpacs,superscriptaddress,prl,citeautoscript,amsmath,amssymb,letterpaper]{revtex4-2}

\usepackage{graphicx}
\usepackage{multirow}
\usepackage{color}
\usepackage{bm}
\usepackage{times}
\usepackage{amsmath,bm,amsfonts}
\usepackage{dcolumn}
\usepackage{graphicx}
\usepackage{latexsym}
\usepackage{hhline}
\usepackage{braket}
\usepackage{bbold}

\def\d{\partial}

\begin{document}
\title{
Topological Enhancement of Nonlinear Transports in Unconventional Point-Node Semimetals
}

%\author{Authors}
\author{Junyeong \surname{Ahn}}
\affiliation{Department of Physics, Harvard University, Cambridge, MA 02138, USA}

\date{\today}

\begin{abstract}
The topological singularity of the Bloch states close to the Fermi level significantly enhances nonlinear electric responses in topological semimetals. Here, we systematically characterize this enhancement for a large class of topological nodal-point fermions, including those with linear, linear-quadratic, and quadratic dispersions. Specifically, we determine the leading power-law dependence of the nonlinear response functions on the chemical potential $\mu$ defined relative to the nodal point. We identify two characteristics that qualitatively improve nonlinear transports compared to those of conventional Dirac and Weyl fermions. First, the type II (over-tilted) spectrum leads to the $\log\mu$ enhancement of nonlinear response functions having zero scaling dimension with respect to $\mu$, which is not seen in a type-I (moderately or not tilted) spectrum. Second, the anisotropic linear-quadratic dispersion increases the power of small-$\mu$ divergence for the nonlinear response tensors along the linearly dispersing direction.
Our work reveals new experimental signatures of unconventional nodal points in topological semimetals as well as provides a guiding principle for giant nonlinear electric responses.
\end{abstract}

\maketitle

{\it Introduction.---}
Nonlinear electromagnetic responses of quantum materials are recently getting attention for both scientific and technological applications~\cite{tokura2018nonreciprocal,liu2020semimetals,dai2023recent,du2021nonlinear,orenstein2021topology,ma2022photocurrent}.
They probe the symmetry of materials that are not easily seen through linear responses~\cite{orenstein2021topology,ma2022photocurrent,zhao2017global,xu2020spontaneous} and lead to various functional properties that may be used in future technologies, such as energy harvesting, wireless communications, and ultrafast information processing~\cite{tokura2018nonreciprocal,liu2020semimetals,dai2023recent,ghimire2019high,isobe2020high,onishi2022high,ahn2020low}.
%For technological applications, in particular, it is an important issue to find material platforms that shows large nonlinear responses for the desired purpose.
%Topological semimetals have nice properties in this regard.
%Conventional topological semimetals are zero-gap semiconductors where conduction and valence bands touch at discrete points or lines in momentum space.

Topological singularities formed by band crossing in topological semimetals are potential sources of large nonlinear responses because of the intimate relation between electromagnetic responses and quantum geometry and topology~\cite{deyo2009semiclassical,moore2010confinement,hosur2011circular,gao2014field,sodemann2015quantum,morimoto2016topological,de2017quantized,holder2020consequences,watanabe2021chiral,ahn2020low,ahn2022riemannian}.
%Because of the gapless spectrum, they can effectively respond to photons having small energies corresponding to the band gap regime of normal semiconductors such as silicon.
%Meanwhile, like a semiconductor, they respond strongly to light beyond the gigahertz regime where metals become inefficient.
%Therefore, nonlinear responses in topological semimetals are promising for applications to terahertz technology~\cite{liu2020semimetals,ahn2020low}.
%Weyl and Dirac semimetals~\cite{armitage2018weyl,murakami2007phase,wan2011topological,burkov2011weyl,young2012dirac,wang2012dirac,wang2013three}, which host point nodes where two energy levels touch with linear dispersions, are conventional topological semimetals.
Recently, topological semimetals have been generalized beyond conventional Weyl and Dirac semimetals~\cite{armitage2018weyl,murakami2007phase,wan2011topological,burkov2011weyl,young2012dirac,wang2012dirac,wang2013three} in various directions to include
the overtilted so-called type-II spectrum~\cite{soluyanov2015type,li2021type},
quadratic or higher-order dispersions~\cite{fang2012multi,yang2014classification,zhang2020twofold},
crossing of more than two bands~\cite{kennett2011birefringent,wieder2016double,bradlyn2016beyond},
line and surface nodes~\cite{kim2015dirac,fang2015new,wu2018nodal}.
%The low-energy excitations around these nodes are unconventional in the sense that they are distinct from relativistic Weyl and Dirac fermions as well as non-relativistic free electrons.
These new types of topological singularities are promising platforms that may have unique nonlinear response properties~\cite{yang2017divergent,ahn2020low}.

In this paper, we investigate topological enhancements of nonlinear electric response by unconventional topological point nodes.
We focus on the responses from Fermi surfaces and study how they vary as the chemical potential $\mu$ approaches a point node.
In contrast to nodal lines or surfaces, which are extended over a finite energy window in general~\cite{ahn2017electrodynamics}, a nodal point resides at a single energy level.
Therefore, the full topological singularity of a node is probed as $\mu$ is tuned close to its energy level.
We consider low-energy models of single nodal points whose relative momentum coordinates have fixed scaling dimensions with respect to the energy.
These include linearly dispersing nodal-point fermions, double Weyl and Dirac fermions having anisotropic linear-quadratic dispersions, and isotropic quadratic band crossing.
We find qualitatively different behaviors in normal (type I) and overtilted  (type II) cases when the scaling dimension of a response function is zero.
The type-II spectrum shows logarithmically enhanced responses at small $\mu$.
Moreover, we show that an anisotropic scaling in a double Weyl or Dirac fermion leads to significantly increased nonlinear responses along the linearly dispersing direction for both types I and II.
Our work shows that unconventional nodal fermions are sources of giant nonlinear electric responses.
Furthermore, our results provide new experimental signatures of unconventional nodal fermions in nonlinear responses.

\begin{table}[b!]
\begin{tabular}{c|cc}
Response	type		& Nonlinear Drude weight & Berry curvature multipole \\
\hhline{=|==}
Expression&Eq.~\eqref{eq:Drude}&Eq.~\eqref{eq:Berry}\\
$T$ or $CPT$ & $(-1)^{N+1}$ & $(-1)^{N}$\\
$PT$ & $+1$ & $-1$\\
$P$ or $C$ & $(-1)^{N+1}$ & $(-1)^{N+1}$\\
$\mu$ dependence &$O(\mu^{\Delta_{\cal D}})$ or $O(\log^s\mu)$	&$O(\mu^{\Delta_{\cal D}-1})$ or $O(\log^s\mu)$
\end{tabular}
\caption{
Properties of nonlinear electric responses near a nodal point.
The third-to-fifth rows show whether the nonlinear Drude weight or the Berry curvature multipole changes sign under the action of charge conjugation $C$, spatial inversion $P$, time reversal $T$, or their combination.
In the last row, $\Delta_{\cal D}$ is the scaling dimension of the $N$th-order Drude weight ${\cal D}_{a;a_1\hdots a_N}$ with respect to the chemical potential $\mu$ away from the nodal point.
For linear dispersion along all directions, $\Delta_{\cal D}=d-N$, where $d$ is the spatial dimension, does not depend on the direction of the tensor components.
In general, though, the scaling dimension $\Delta_{{\cal D}_{a;a_1\hdots a_N}}$ depends on the direction of each tensor component (see, e.g., Table~\ref{tab:Dirac-Weyl2}).
When the scaling dimension is zero, the logarithm may appear as the leading term.
$s=0$ and $1$ represent type I and type II nodal points, respectively.
}
\label{tab:symmetry}
\end{table}

{\it Symmetries of nonlinear electric responses.---}
In our study, we work in the perturbative response regime where the applied electric field is not extremely strong, and we do not include interaction and disorder effects.
The $N$th-order nonlinear responses to electric fields ${\bf E}(\omega_i)$ oscillating with angular frequencies $\omega_i$ are described by $j^a_{(N)}(\omega)=\sum_{a_i,\omega_i}\tilde{\sigma}_{a;a_1\hdots a_N}(\omega;\omega_1,\hdots, \omega_N)E^{a_1}(\omega_1)\hdots E^{a_N}(\omega_N)$, where $\omega=\sum_{i=1}^N\omega_i$.
The nonlinear electric conductivity tensor has two intrinsic Fermi surface contributions in the leading order of $\omega^{-1}$~\cite{sodemann2015quantum,parker2019diagrammatic}:
$\tilde{\sigma}_{a;a_1\hdots a_N}
\propto \omega^{-1}{\cal D}_{a;a_1\hdots a_N}+{\cal B}_{a;a_1\hdots a_N}$, where
\begin{align}
\label{eq:Drude}
{\cal D}_{a;a_1\hdots a_N}
&=\sum_n\int \frac{d^dk}{(2\pi)^d}f_n\d_{a_1}\hdots\d_{a_N}\d_a\epsilon_n
\end{align}
is the generalized Drude weight, and
\begin{align}
\label{eq:Berry}
{\cal B}_{a;a_1\hdots a_N}
&=\sum_{n}\int \frac{d^dk}{(2\pi)^d}f_n\d_{a_2}\hdots\d_{a_{N}}F^{aa_1}_n
\end{align}
is the Berry curvature multipole ($N$-pole), which satisfies ${\cal B}_{a;a_1\hdots a_N}=-{\cal B}_{a_1;a\hdots a_N}$ under the exchange of the first two indices.
Here, $\epsilon_n$ is the energy of band $n$, $f_n$ is the corresponding Fermi-Dirac distribution, $\d_a=\d/\d k_a$, and $F^{ab}_n=\d_a\braket{u_n|\d_{b}|u_n}-\d_b\braket{u_n|\d_{a}|u_n}$ is the Berry curvature.
The DC limit can be taken after replacing $\omega_i$ with $\omega_i+i\tau^{-1}$ with a phenomenological relaxation time $\tau$~\cite{passos2018nonlinear}.

Let us investigate some important symmetry properties of  ${\cal D}$ and ${\cal B}$.
They both transform as
${\cal D}_{a;a_1\hdots a_N}
\rightarrow (-1)^{N+1}{\cal D}_{a;a_1\hdots a_N}$
and
${\cal B}_{a;a_1\hdots a_N}
\rightarrow (-1)^{N+1}{\cal B}_{a;a_1\hdots a_N}$
under spatial inversion $P:{\bf r}\rightarrow {\bf -r}$.
One way to see this is to note that the current and electric fields all flip the direction under inversion.
On the other hand, they transform oppositely under time reversal $T:t\rightarrow -t$ because the Berry curvature is odd under time reversal.
Under spacetime inversion $PT:(t,{\bf r})\rightarrow (-t,-{\bf r})$,
$PT:
{\cal D}_{a;a_1\hdots a_N}
\rightarrow {\cal D}_{a;a_1\hdots a_N},
\quad
{\cal B}_{a;a_1\hdots a_N}
\rightarrow -{\cal B}_{a;a_1\hdots a_N}$.
We thus see that ${\cal B}_{a;a_1\hdots a_N}=0$ in Dirac or nodal line semimetals protected by $PT$ symmetry.
Since ${\cal D}$ and ${\cal B}$ transform in the same way under $P$ and oppositely under $PT$, they transform oppositely under $T$.

In our analysis of the low-energy response properties near a nodal point, $CPT$ symmetry also plays an important role, where $C$ is the charge conjugation operator.
The $CPT$ theorem in higher-energy physics implies that an electronic system also has a  $CPT$ symmetry when the system has linear dispersion at the chemical potential.
In fact, $CPT$ symmetry always exists for a ${\bf k}$-linear Hamiltonian, because the symmetry condition $(CPT)H({\bf k})(CPT)^{-1}=-H(-{\bf k})$ follows if we take $CPT=1$ and use $H({\bf k})=-H(-{\bf k})$.
Therefore, the behavior of the nonlinear response when the chemical potential approaches a linearly dispersing nodal point is constrained by the emergent $CPT$ symmetry~\cite{ahn2020low}.
The action of the charge conjugation on ${\cal D}$ and ${\cal B}$ give the factor $(-1)^{N+1}$.
This can be understood from the transformation properties of the current density and electric field, both of which changes sign under charge conjugation, and the definition of the nonlinear electric conductivity.
Combining the action of $C$ and $PT$, we have $CPT:{\cal D}_{a;a_1\hdots a_N} \rightarrow (-1)^{N+1}{\cal D}_{a;a_1\hdots a_N}$ and ${\cal B}_{a;a_1\hdots a_N} \rightarrow (-1)^{N}{\cal B}_{a;a_1\hdots a_N}$.
These symmetry constraints show that the nonlinear Drude (Berry curvature) response of a linearly dispersing system vanishes at the even (odd) order when the chemical potential is exactly at the nodal point.
The $C$, $P$, and $T$ symmetry transformation properties are summarized in Table~\ref{tab:symmetry}.

{\it Scaling relations.---}
We are interested in the case where symmetries allow a nonzero response when the chemical potential is at a nodal point.
Let us first estimate the size of the response with dimensional analysis.
The generalized Drude weight has the dimension of $[{\cal D}]=[k_1]\cdots [k_d] [k_{a_1}]^{-1}\cdots[k_{a_N}]^{-1}[k_{a}]^{-1}[E]$, and $[{\cal B}]=[{\cal D}][E]^{-1}$.
We suppose that the low-energy degrees of freedom near the nodal point follow scaling relation betweens the Fermi momentum and chemical potential
\begin{align}
\label{eq:scaling-k}
k_{a}\propto \mu^{\Delta_{k_a}},
\end{align}
where the scaling exponent, in general, depends on the momentum component.
This scaling relation holds for the Hamiltonians in Eqs.~\eqref{eq:W-H} and~\eqref{eq:DW-H} below.

Then, we naturally expect ${\cal D}_{a;a_1\hdots a_N}\propto \mu^{\Delta_{{\cal D}_{a;a_1\hdots a_N}}}$ and ${\cal B}_{a;a_1\hdots a_N}\propto \mu^{\Delta_{{\cal B}_{a;a_1\hdots a_N}}}$ when they does not vanish by symmetry as $\mu\rightarrow 0$, with the scaling exponent determined by their dimensions:
\begin{align}
\Delta_{{\cal D}_{a;a_1\hdots a_N}}
&=\sum_{i=1}^d\Delta_{k_i}-\sum_{j=0}^N\Delta_{k_{a_j}}+1,\notag\\
\Delta_{{\cal B}_{a;a_1\hdots a_N}}
&=\Delta_{{\cal D}_{a;a_1\hdots a_N}}-1,
\end{align}
where $a_0=a$.
When the scaling dimension of ${\cal D}$ (${\cal B}$) is zero, one expects that ${\cal D}$ (${\cal B}$) approaches a constant as $\mu\rightarrow 0$.
This is the case for a type-I nodal point, where the density of states vanishes at $\mu=0$.
On the other hand, however, logarithmic dependence may appear for a type-II nodal point, which has a finite density of states at $\mu=0$.
Therefore, we have the following divergent behavior at small $\mu$.
\begin{align}
\label{eq:scaling}
{\cal O}_{a;a_1\hdots a_N}
&= 
\begin{cases}
O(\mu^{\Delta_{{\cal O}_{a;a_1\hdots a_N}}})&\text{ when }\Delta_{{\cal O}_{a;a_1\hdots a_N}}<0,\\
O(\log^s\mu)&\text{ when }\Delta_{{\cal O}_{a;a_1\hdots a_N}}=0,
\end{cases}
\end{align}
where ${\cal O}={\cal D}$ or ${\cal B}$, and $s=0$ for type I and $s=1$ for type II. 
When the scaling dimension is isotropic, i.e., $\Delta_{k_a}=\Delta_{k_{a_i}}=\Delta_k$ for all $i=1,\hdots,d$, $\Delta_{\cal B}=(d-N-1)\Delta_k$ and $\Delta_{\cal D}=\Delta_{\cal B}+1$ is minimized when $\Delta_k$ is the largest and $d$ is the smallest for $d-N-1<0$.
Therefore, linear dispersion and low dimensions are the requirements for strong nonlinear responses at small $\mu$ for isotropic scaling dimensions (see Ref.~\cite{ahn2020low} for a related discussion for nonlinear optical responses at small frequencies).

The appearance of the logarithm for type II manifests that the Fermi surface is extended over a large range of momentum scales, from small to large.
The response function is given by integration over the momentum scales, such that it contains $\int^{k_{\Lambda}}_{k_{\mu}} dk/k=\log (k_{\Lambda}/k_{\mu})$, where $k_{\mu}$ and $k_{\Lambda}$ are lower and upper cutoffs, respectively, for a momentum.
This is in contrast to type I, where there exists a momentum scale that determines the size of the Fermi surface, and no logarithmic dependence appears.

%%%%%%%%%%%%%%%%%%%%%%%%%%%%%%%%%%%%%%%%   FIGURE   %%%%%%%%%%%%%%%%%%%%%%%%%%%%%%%%%%%%%%%%%%%%%%%%%%%%%%
\begin{figure}[t]
\includegraphics[width=0.48\textwidth]{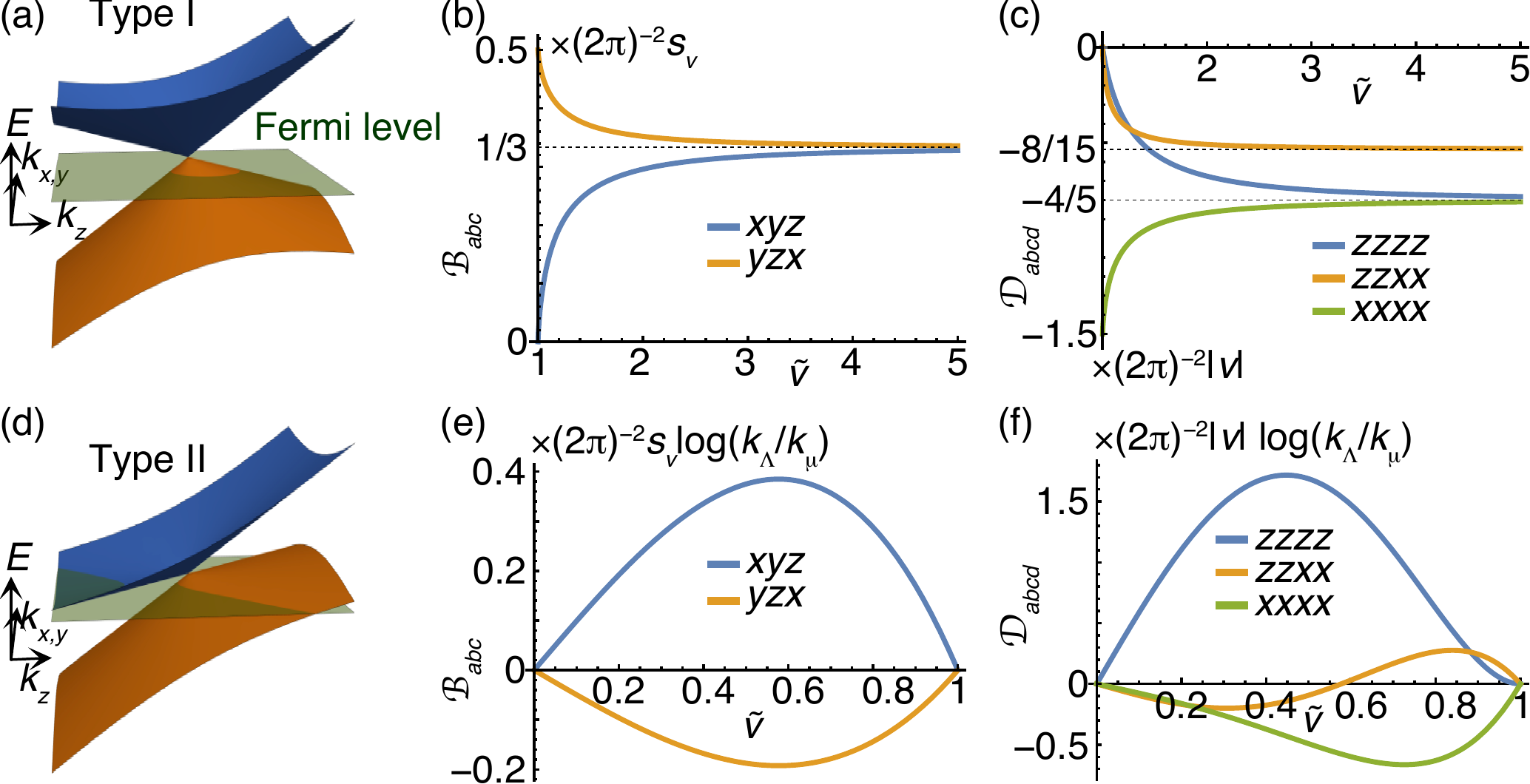}
\caption{
Berry curvature dipole ${\cal B}_{abc}$ and third-order Drude weight ${\cal D}_{abcd}$ of a linearly dispersing Weyl fermion.
(a-c), Type I Weyl fermion.
(a) A typical type-I energy spectrum.
The green sheet represents the Fermi level.
(b,c) ${\cal B}_{abc}$ and ${\cal D}_{abcd}$ as a function of $\tilde{v}=v/u>1$, which is inversely proportional to the tilting of the Weyl cone.
(d-f), Type II Weyl fermion.
(d), A typical type-II energy spectrum.
(e,f) ${\cal B}_{abc}$ and ${\cal D}_{abcd}$ as a function of $0<\tilde{v}<1$.
$k_{\mu}$ and $k_{\Lambda}$ are respectively lower and upper cutoffs for the momentum, where $k_{\mu}\propto \mu$.
}
\label{fig:Weyl}
\end{figure}
%%%%%%%%%%%%%%%%%%%%%%%%%%%%%%%%%%%%%%%%   FIGURE   %%%%%%%%%%%%%%%%%%%%%%%%%%%%%%%%%%%%%%%%%%%%%%%%%%%%%%

We demonstrate the result of our scaling analysis with two models in three dimensions: a Weyl fermion model and a double Weyl fermion model.
Since a (double) Dirac fermion is two copies of (double) Weyl fermion on top of each other with compensating Berry curvature, their scaling relation is the same for the nonlinear Drude response.

{\it Weyl fermion in three dimensions.---}
We first consider a three-dimensional Weyl fermion tilted along the $z$ direction.
\begin{align}
\label{eq:W-H}
H_{W}
=-\mu+u k_z +v(k_x\sigma_x+k_y\sigma_y+k_z\sigma_z),
\end{align}
where $\sigma_{x,y,z}$ are Pauli matrices.
It is type I when $|v/u|>1$ and type II when $|v/u|<1$.
For type I (type II), the density of states vanishes (remains finite) when the chemical potential is at the nodal point, i.e., $\mu=0$.
When $\mu=0$, the Hamiltonian is linear in ${\bf k}$, and it thus have $CPT$ symmetry under $CPT=1$.

Let us consider the behavior of second- and third-order responses as $\mu\rightarrow 0$.
Since $CPT$ symmetry imposes ${\cal D}_{abc}={\cal B}_{abcd}=0$ at $\mu=0$, the second-order response at $\mu=0$ is due to ${\cal B}_{abc}$ while the third-order response is due to  ${\cal D}_{abcd}$.
The scaling dimensions of ${\cal B}_{abc}$ and ${\cal D}_{abcd}$ are both zero because $\Delta_{k_x}=\Delta_{k_y}=\Delta_{k_z}=1$ by the linear dispersion along all directions.

${\cal B}_{abc}$ has two independent nonvanishing components: ${\cal B}_{xyz}$ and ${\cal B}_{yzx}$.
For type I, they are finite as $\mu\rightarrow 0$, and their expressions were derived in Ref.~\cite{matsyshyn2019nonlinear}:
${\cal B}_{xyz}=-(2\pi)^{-2}s_v(\tilde{v}^2-1)\left(1-\tilde{v}\coth^{-1}\tilde{v}\right)$, and
${\cal B}_{yzx}=2^{-1}(2\pi)^{-2}s_v\left(\tilde{v}^2-\tilde{v}(\tilde{v}^2-1)\coth^{-1}\tilde{v}\right)$
for $|\tilde{v}|>1$, where $s_v=v/|v|$ and $\tilde{v}=v/u$.
Both components saturate to $1/3$ times $1/(2\pi)^2s_v$ in the limit of zero tilting $|\tilde{v}|\rightarrow \infty$, as shown in Fig.~\ref{fig:Weyl}(a).
A nonzero tilting leads to an anisotropy of the Berry curvature at the Fermi surface such that ${\cal B}_{xyz}\ne {\cal B}_{yzx}$.
Since the sum ${\cal B}_{xyz}+{\cal B}_{yzx}+{\cal B}_{zxy}={\cal B}_{xyz}+2{\cal B}_{yzx}$ measures the Berry flux across the Fermi surface, so it has a nonzero quantized value independent of $\tilde{v}$, as long as $|\tilde{v}|>1$.
For type II,
${\cal B}_{xyz}=(2\pi)^{-2}s_v|\tilde{v}|(1-\tilde{v}^2)\log (k_{\Lambda}/k_{\mu})$ and ${\cal B}_{yzx}=-2^{-1}(2\pi)^{-2}s_v\tilde{v}|(1-\tilde{v}^2)\log (k_{\Lambda}/k_{\mu})$,
diverge logarithmically as the small-momentum cutoff $k_{\mu}$ approaches zero.
In contrast to the case of type I, ${\cal B}_{xyz}+2{\cal B}_{yzx}=0$ at $\mu=0$ for type II because the Berry flux across the Fermi surface is zero.

\begin{table}[t!]
\begin{tabular}{c|cccc}
Response functions	& ${\cal D}_{abc}$ 	&  ${\cal B}_{abc}$ 	& ${\cal D}_{abcd}$ 	& ${\cal B}_{abcd}$ \\
\hhline{=|====}
2D Dirac			& $0$ 			& $0$ 			& $O(\mu^{-1})$	& $0$\\
3D Weyl 			& $0$			& $O(\log^s\mu)$	& $O(\log^s\mu)$ 	& $0$\\
3D Dirac 			& $0$			& $0$  			& $O(\log^s\mu)$ 	& $0$
\end{tabular}
\caption{
The leading small-$\mu$ divergence of the second- and third-order electric responses of linear Weyl and Dirac fermions in two and three dimensions.
``$0$" indicates the absence of divergence, converging to a finite value or zero.
${\cal D}_{abc}={\cal B}_{abcd}\rightarrow 0$ as $\mu\rightarrow 0$ by emergent $CPT$ symmetry, and ${\cal B}=0$ for Dirac points by $PT$ symmetry.
$s=0$ and $1$ correspond to type-I and type-II dispersions, respectively.
}
\label{tab:Dirac-Weyl}
\end{table}

The third-order Drude function has three independent nonvanishing components: ${\cal D}_{zzzz}$, ${\cal D}_{zzxx}$, ${\cal D}_{xxxx}$.
As they have scaling dimension zero, they are finite as $\mu\rightarrow 0$ for type I [Fig.~\ref{fig:Weyl}(c)].
${\cal D}_{zzzz}=-2(2\pi)^{-2}|v|(\tilde{v}^2-1)\left(3\tilde{v}^2-2-3\tilde{v} (\tilde{v}^2-1) \coth^{-1}\tilde{v}\right)$,
${\cal D}_{zzxx}=2(2\pi)^{-2}|v|(\tilde{v}^2-1)\left(3\tilde{v}^2-(3\tilde{v}^2-1)\tilde{v}\coth^{-1}\tilde{v}\right)$, and
${\cal D}_{xxxx}=-(3/4)(2\pi)^{-2}|v|\left(3\tilde{v}^4-\tilde{v}^2-(3\tilde{v}^4-2\tilde{v}^2-1)\tilde{v}\coth^{-1}\tilde{v}\right)$.
for $|\tilde{v}|>1$.
They converge to $-4/5$, $-8/15$, and $-4/5$, respectively, in unit of $(2\pi)^{-2}|v|$ as $|\tilde{v}|\rightarrow \infty$ and vanish as $|\tilde{v}|\rightarrow 1$.
For type II, logarithmic divergence appears as $\mu\rightarrow 0$.
${\cal D}_{zzzz}=6(2\pi)^{-2}|v\tilde{v}|(1-\tilde{v}^2)^2\log (k_{\Lambda}/k_{\mu})$,
${\cal D}_{zzxx}=(2\pi)^{-2}|v\tilde{v}|(1-\tilde{v}^2)(3\tilde{v}^2-1)\log (k_{\Lambda}/k_{\mu})$, and
${\cal D}_{xxxx}=-(3/4)(2\pi)^{-2}|v\tilde{v}|(1-\tilde{v}^2)(3\tilde{v}^2+1)\log (k_{\Lambda}/k_{\mu})$.
These all vanish at $\tilde{v}=0$ and $\tilde{v}=1$ and reach the maximum in between [Fig.~\ref{fig:Weyl}(d)].
While the third-order Drude weights are negative for type I, they may have positive values for type II.

Table~\ref{tab:Dirac-Weyl} summarizes the leading small-$\mu$ divergence of nonlinear response functions for a Weyl fermion.
The case of two- and three-dimensional Dirac fermions is also shown for comparison.

%%%%%%%%%%%%%%%%%%%%%%%%%%%%%%%%%%%%%%%%   FIGURE   %%%%%%%%%%%%%%%%%%%%%%%%%%%%%%%%%%%%%%%%%%%%%%%%%%%%%%
\begin{figure}[t]
\includegraphics[width=0.48\textwidth]{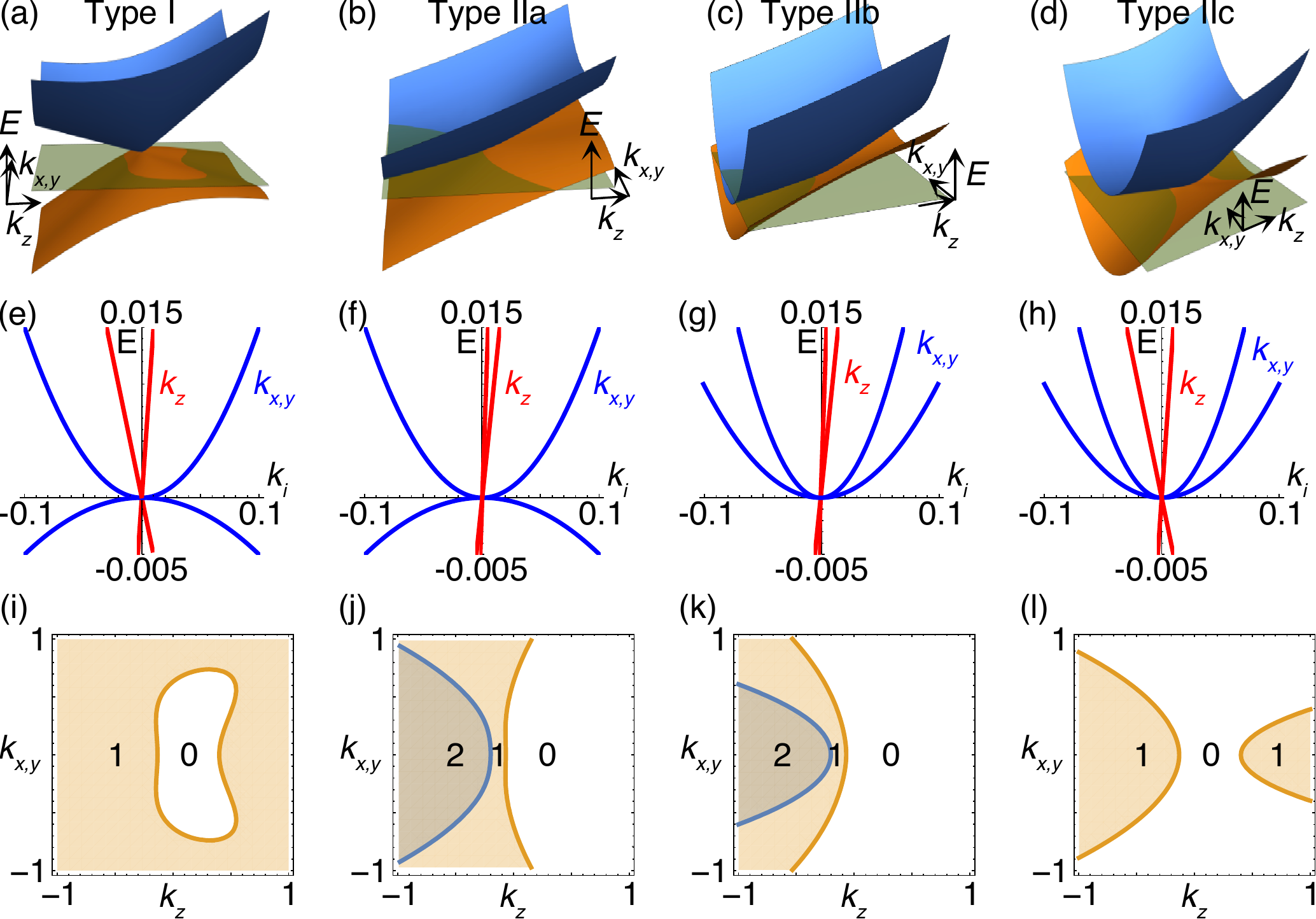}
\caption{
Spectral properties of a double Weyl fermion.
(a-d), Four types of energy spectra.
The green sheet represents the Fermi level.
(e-h), Dispersion relation along lines crossing ${\bf k}=(0,0,0)$.
The red and blue curves are the energy spectra along $k_x$ and $k_{\parallel}$ directions, respectively, where $k_{\parallel}$ is in the $k_x$-$k_y$ plane.
(i-l), Electron occupancy and the Fermi surface.
The numbers $0$, $1$, and $2$ show the occupancy of the region.
The blue and gold lines represent the Fermi surfaces by the upper (blue) and lower (gold) part of the Dirac cone in {\bf a-d}.
All figures are plotted with the model in Eq.~\eqref{eq:DW-H}, where $v=1$, $B=1$, and $\mu=-0.2$.
(a,e,i) $u=0.5$ and $A=0.5$.
(b,f,j) $u=2$ and $A=0.5$.
(c,g,k) $u=2$ and $A=2$.
(d,h,l) $u=2$ and $A=0.5$.
}
\label{fig:DW}
\end{figure}
%%%%%%%%%%%%%%%%%%%%%%%%%%%%%%%%%%%%%%%%   FIGURE   %%%%%%%%%%%%%%%%%%%%%%%%%%%%%%%%%%%%%%%%%%%%%%%%%%%%%%

{\it Double Weyl fermion in three dimensions.---}
When there is a $C_{n=3,4,6}$ symmetry, double Weyl fermions can be stabilized along the rotation-invariant axis in the three-dimensional Brillouin zone.
We take $\hat{z}$ as the rotation axis and assume symmetry under $C_{nz}=\exp(-i\pi n\sigma_z)$ with $n=3$ ,$4$, or $6$.
The low-energy effective Hamiltonian of a double Weyl point in the lowest order in relative momentum takes the form
\begin{align}
\label{eq:DW-H}
H_{\rm DW}
=&-\mu+u k_z+A(k_x^2+k_y^2)\notag\\
&+B[(k_x^2-k_y^2)\sigma_x+2k_xk_y\sigma_y]+vk_z\sigma_z.
\end{align}
As in the case of a linearly dispersing Weyl fermion, we distinguish two types of dispersion depending on whether the density of states is zero or finite at $\mu=0$.
A double Weyl fermion is type I when $A^2<(1-u^2/v^2)B^2$ and type II when $A^2>(1-u^2/v^2)B^2$.
The type-II dispersion can appear because of two different origins: over-tilting along the $z$ direction $|u|>|v|$ and large in-plane curvature $A^2>(1-u^2/v^2)B^2$ independent of tilting along $z$.
While the latter case was originally termed type III~\cite{li2021type}, we prefer to call it also type II because it shows no qualitative difference in the small-$\mu$ behavior.
We define type IIa, IIb, and IIc as the cases with $|u|>|v|$ and $A^2<(1-u^2/v^2)B^2$, $|u|>|v|$ and $A^2>(1-u^2/v^2)B^2$, and $|u|<|v|$ and $A^2>(1-u^2/v^2)B^2$, respectively [Fig.~\ref{fig:DW}].
In contrast to the case of a linear Weyl fermion, a double Weyl fermion does not have $CPT$ symmetry at $\mu=0$, as long as $A\ne 0$.

We consider second- and third-order responses for $\mu\rightarrow 0$.
As there is no $CPT$ symmetry for a double Weyl fermion, ${\cal D}_{abc}$ and ${\cal B}_{abcd}$ as well as ${\cal D}_{abcd}$ and ${\cal B}_{abc}$ can remain nonzero at $\mu=0$.
We see above that the analytic expressions are already complicated for a single Weyl fermion.
Here, instead of presenting the full analytic expressions, we show the scaling relations numerically.

\begin{figure}[t!]
\includegraphics[width=0.5\textwidth]{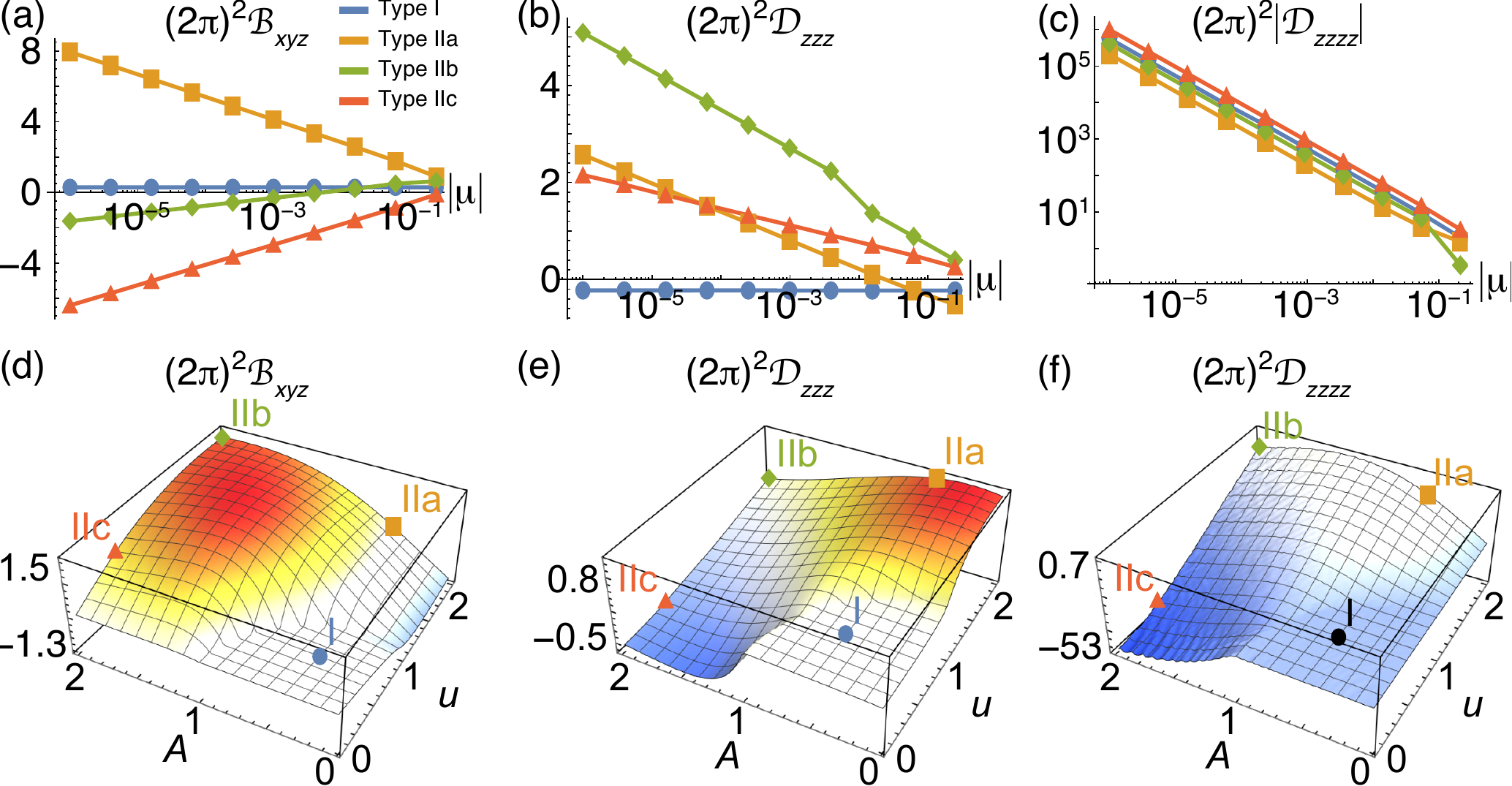}
\caption{
Berry curvature dipole and nonlinear Drude weights of a double Weyl fermion.
$v=B=1$ are fixed for all plots, and $\mu<0$.
Only components (a,d) $(2\pi)^2{\cal B}_{xyz}$,  (b,e) $(2\pi)^2{\cal D}_{zzz}$,  (c,f) $(2\pi)^2{\cal D}_{zzzz}$ are shown, but other nonvanishing components also exist (see Table~\ref{tab:Dirac-Weyl2} and Supplemental Material~\cite{supp}).
(a-c) Chemical potential dependence.
The horizontal axes are shown in the logarithmic scale.
Response functions are calculated at $(u,A)=(0,0)$, $(2,0.5)$, $(2,2)$, $(0.5,2)$,  which are type I, IIa, IIb, and IIc, respectively.
We take the absolute value $|{\cal D}_{zzzz}|$ for log-log plots in (c).
${\cal D}_{zzzz}>0$ for type IIb except at $\mu=-0.06$, where ${\cal D}_{zzzz}<0$, and ${\cal D}_{zzzz}<0$ for type I, IIa, and IIc.
(d-f) $u$ and $A$ dependence at a fixed chemical potential $\mu=-0.01$.
For all calculations, the momentum is cut off by $|k_z|,\sqrt{k_x^2+k_y^2}\le 1$.
}
\label{fig:DW-response}
\end{figure}

Figures~\ref{fig:DW-response}(a-c) show the scaling of selected nonvanishing nonlinear response functions.
${\cal B}_{xyz}$ and ${\cal D}_{zzz}$ are almost constant in $\mu$ for type I and have the $\log\mu$ dependencies for type II, as expected from the scaling analysis Eq.~\eqref{eq:scaling}.
The magnitude of ${\cal D}_{zzzz}$ follows the $\mu^{-1}$ scaling for both type I and II.
The relative enhancement of ${\cal D}_{zzzz}$ compared to the case of a linear Weyl fermion at small $\mu$ is attributed to the increased density of states.
In the case of ${\cal B}_{xyz}$, the effect of the increased density of states is compensated by the weaker geometric singularity of quantum states along $k_{x,y}$.
The $u$ and $A$ dependence of the response functions are shown in Figs.~\ref{fig:DW-response}(d-f) with $v=B=1$ and $\mu=-0.01$.
fermion.
We show the numerical data for other components in Supplemental Materials~\cite{supp} and summarize the full small-$\mu$ divergence in Table~\ref{tab:Dirac-Weyl2}.

%%%%%%%%%%%%%%%%%%%%%%%%%%%%%%%%%%%%%%%%   FIGURE   %%%%%%%%%%%%%%%%%%%%%%%%%%%%%%%%%%%%%%%%%%%%%%%%%%%%%%
%%%%%%%%%%%%%%%%%%%%%%%%%%%%%%%%%%%%%%%%   FIGURE   %%%%%%%%%%%%%%%%%%%%%%%%%%%%%%%%%%%%%%%%%%%%%%%%%%%%%%

\begin{table*}[t]
\begin{tabular}{c|ccccccccc}
Response functions		& ${\cal D}_{zzz}$ 	& ${\cal D}_{zxx}$ 	& ${\cal B}_{xyz},{\cal B}_{yzx}$ 	& ${\cal D}_{zzzz}$ 	& ${\cal D}_{zzxx}$	& ${\cal D}_{xxxx},{\cal D}_{xxyy}$	&${\cal B}_{zxyz},{\cal B}_{xyzz}$\\
\hhline{=|=========}
3D double Weyl 		& $O(\log^s\mu)$	& $0$			& $O(\log^s\mu)$	& $O(\mu^{-1})$ 	& $O(\log^s\mu)$		& $0$			& $O(\mu^{-1})$\\
3D double Dirac 		& $O(\log^s\mu)$	& $0$			& $0$			& $O(\mu^{-1})$ 	& $O(\log^s\mu)$		& $0$			& $0$
\end{tabular}
\caption{
The leading small-$\mu$ divergence of the second- and third-order electric responses of double Weyl and Dirac fermions in three dimensions.
The double Weyl fermion is described by the low-energy effective Hamiltonian $H_{\rm DW}$ in Eq.~\eqref{eq:DW-H}, and the double Dirac fermion is described by $\tau_z\otimes H_{\rm DW}$, where $\tau_z$ is a Pauli matrix.
``$0$" indicates the absence of divergence, converging to a finite value or zero.
${\cal B}=0$ for Dirac points by $PT$ symmetry.
$s=0$ and $1$ correspond to type-I and type-II dispersions, respectively.
}
\label{tab:Dirac-Weyl2}
\end{table*}

{\it Discussion.---}
While we focus on Weyl fermions with linear or quadratic dispersions above, a three-dimensional triple Weyl fermion, having a linear-cubic-cubic dispersion, can also be stabilized by rotational symmetries.
When applying our analysis to such a triple Weyl fermion, one might naively expect even more enhanced responses such as ${\cal D}_{zzz}=O(\mu^{-1/3})$ and ${\cal D}_{zzzz}=O(\mu^{-4/3})$ by assuming the cubic scaling relation along $k_x$ and $k_y$: $\Delta_{k_{x,y}}=1/3$.
However, this scaling relation does not hold in general because the Hamiltonian has a quadratic term $\propto (k_x^2+k_y^2)$ as an overall energy shift, which dominates the dispersion at small ${\bf k}$.
The same argument applies to a quartic Weyl fermion~\cite{zhang2020twofold}, which appears only in systems with negligible spin-orbit coupling.

There exist many potential material platforms~\cite{lv2021experimental} for giant nonlinear transports, including type-II Weyl semimetals Mo$_{1-x}$W$_x$Te$_2$~\cite{belopolski2016discovery}, LaAlGe~\cite{xu2017discovery}, TaIrTe$_4$~\cite{haubold2017experimental,belopolski2017signatures} and double Weyl semimetals SrSi$_2$~\cite{huang2016new} and (TaSe$_4$)$_2$I~\cite{li2021type}.
One way to see the manifestation of the divergently enhanced nonlinear conductivity is to vary the temperature $T$.
Most nodal-point semimetals, with the exception of graphene, have nodal points away from the Fermi level.
Since the chemical potential depends on the temperature, it may be put at the energy level of nodal points, leading to greatly enhanced nonlinear responses.
This may serve as a signature of nodal points in the band structure.
A complication may arise because the nonlinear conductivity depends on transport lifetime $\tau$, which depends on the chemical potential and temperature. 
Nevertheless, we expect that it does not lead to a qualitative difference.
$\tau(T)$ is expected to continuously vary with $T$ when the chemical potential $\mu(T)$ crosses a nodal point because the nodal point has a vanishing density of states.
Another source of complication is coming from interaction effects that modify the scaling dimension~\cite{isobe2016coulomb,huang2017renormalization,lee2018renormalization}.
We leave the analysis of the interaction-induced renormalization for future study.

Lastly, we note that divergent behavior stops at low-energy cutoff provided by the breakdown of perturbative theory or thermal excitations~\cite{supp} or by interaction and disorder effects~\cite{morimoto2016topological,matsyshyn2021rabi}, such that no physical divergence occurs at $\mu=0$.

\begin{acknowledgments}
We appreciate Soohyun Cho for helpful discussions.
This work was supported by the Center for Advancement of Topological Semimetals, an Energy Frontier Research Center funded by the U.S. Department of Energy Office of Science, Office of Basic Energy Sciences, through the Ames Laboratory under contract No. DE-AC02-07CH11358.
\end{acknowledgments}

\clearpage
\newpage

\renewcommand{\thefigure}{S\arabic{figure}}
\renewcommand{\theequation}{S\arabic{equation}}
\renewcommand{\thetable}{S\arabic{table}}
\renewcommand{\thesubsection}{Supplementary Note \arabic{subsection}}

\setcounter{figure}{0} 
\setcounter{equation}{0} 
\setcounter{table}{0} 
\setcounter{section}{0} 

\begin{widetext}

\section{Supplemental Figures}

%%%%%%%%%%%%%%%%%%%%%%%%%%%%%%%%%%%%%%%%   FIGURE   %%%%%%%%%%%%%%%%%%%%%%%%%%%%%%%%%%%%%%%%%%%%%%%%%%%%%%
\begin{figure}[ht!]
\includegraphics[width=\textwidth]{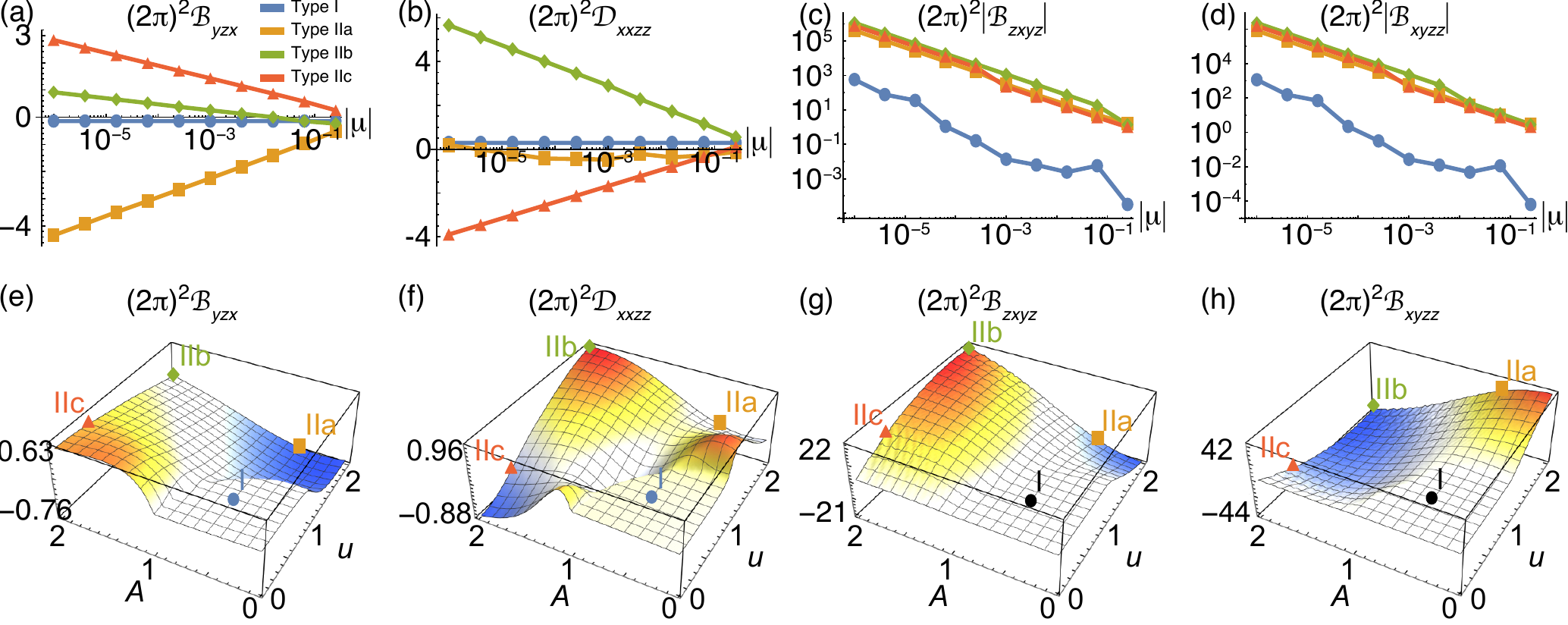}
\caption{
Berry curvature multipole moments and nonlinear Drude weights of a double Weyl fermion.
Here we show only the components not shown in the main text.
We use the same model parameters as in the main text: $v=B=1$ are fixed for all plots, and $\mu<0$.
(a-d) Chemical potential dependence.
(e-h) $u$ and $A$ dependence at a fixed chemical potential $\mu=-0.01$.
For all calculations, the momentum is cut off by $|k_z|,\sqrt{k_x^2+k_y^2}\le 1$.
}
\label{fig:supp1}
\end{figure}
%%%%%%%%%%%%%%%%%%%%%%%%%%%%%%%%%%%%%%%%   FIGURE   %%%%%%%%%%%%%%%%%%%%%%%%%%%%%%%%%%%%%%%%%%%%%%%%%%%%%%

%%%%%%%%%%%%%%%%%%%%%%%%%%%%%%%%%%%%%%%%   FIGURE   %%%%%%%%%%%%%%%%%%%%%%%%%%%%%%%%%%%%%%%%%%%%%%%%%%%%%%
\begin{figure}[ht!]
\includegraphics[width=0.8\textwidth]{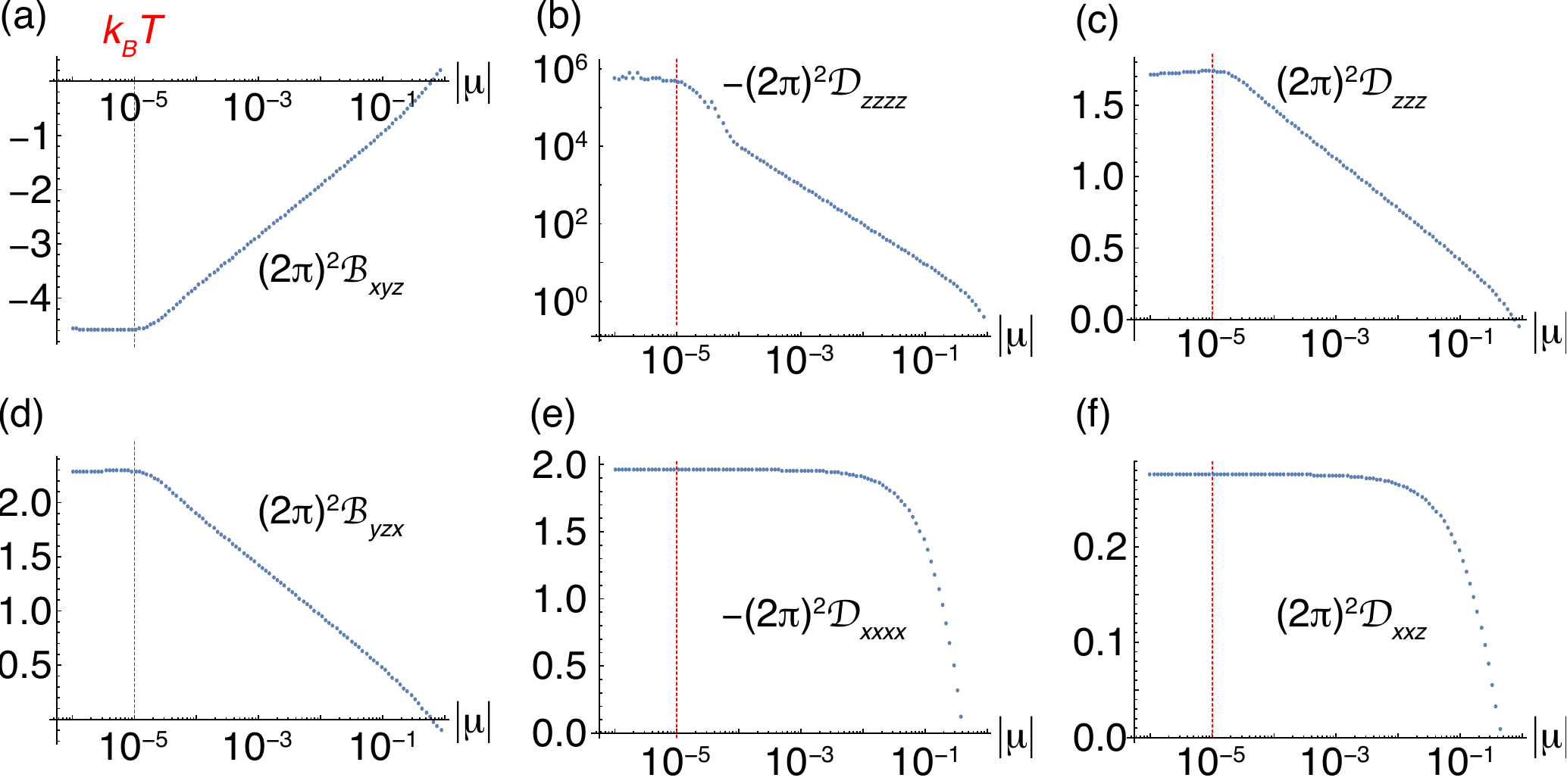}
\caption{
Berry curvature dipole and nonlinear Drude weights of a double Weyl fermion at finite temperature $k_BT=10^{-5}$.
We use $v=B=1$, $u=0.5$, $A=2$, and $\mu<0$.
The temperature scale serves as the low-energy cutoff in (a-d).
The saturations in (e,f) are instead because of the zero scaling dimension.
For all calculations, the momentum is cut off by $|k_z|,\sqrt{k_x^2+k_y^2}\le 1$.
}
\label{fig:supp2}
\end{figure}
%%%%%%%%%%%%%%%%%%%%%%%%%%%%%%%%%%%%%%%%   FIGURE   %%%%%%%%%%%%%%%%%%%%%%%%%%%%%%%%%%%%%%%%%%%%%%%%%%%%%%

\section{Additional discussions}

Let us relate ${\cal D}$ and ${\cal B}$ to nonlinear responses.
Because of the permutation symmetry of the nonlinear conductivity tensor under $(a_i,\omega_i)\leftrightarrow (a_j,\omega_j)$, the physical response depends on a symmetrized tensor $\sigma$ rather than the bare $\tilde{\sigma}$~\cite{boyd2020nonlinear}.
For example, here we consider the DC limit ($\omega_i=0$), where $\sigma_{a;a_1\hdots a_N}=P^{-1}\sum_{P}\tilde{\sigma}_{a;a_1\hdots a_N}$, where $P$ indicates all possible $N!$ permutations of $a_1,\hdots ,a_N$.
While the Drude weight is directly related to the nonlinear conductivity tensor by $\sigma_{a_1\hdots a_N}\propto D_{a_1\hdots a_N}$, the Berry curvature multipole needs to be linearly combined to represent a physical DC response.
In the models we consider, only the difference appears in the second-order electric DC transport through $\sigma_{xyz}=-\sigma_{yxz}=\tau e^2\hbar^{-1}({\cal B}_{xyz}-{\cal B}_{yzx})$ (cf. the sum ${\cal B}_{xyz}+2{\cal B}_{yzx}$ appears in responses to circularly polarized light at optical frequencies~\cite{de2017quantized,matsyshyn2019nonlinear}).
For a linear Weyl fermion, this response increases as tilting increases in type I, but it is maximal at an intermediate tilting in type II [Fig.~\ref{fig:Weyl}].
In the third-order, nonvanishing electric DC conductivity tensors are related to the Berry curvature multipole by $\sigma_{zxxz}={\cal B}_{zxxz}$, $\sigma_{xyzz}=\sigma_{xzyz}={\cal B}_{xyzz}-2{\cal B}_{zxyz}$.
Note that Drude and Berry curvature responses appear in different components of the conductivity tensor in the models we consider.
This separation of responses is attributed to $C_{2z}T$ symmetry~\cite{ahn2020low}.
Since ${\cal B}_{abcd}$ and ${\cal D}_{abcd}$ transform oppositely under $C_{2z}T$, one of them is forbidden by the symmetry.

Although our analysis shows that nonlinear conductivity tensors may diverge as $\mu\rightarrow 0$, they cannot grow indefinitely in reality.
At finite temperature $T$, the thermal energy $k_BT$ serves as a low-energy cutoff scale [Fig.~\ref{fig:supp2}].
Other cutoff scales come into play in the zero temperature limit.
Since the expression for the second-order nonlinear Hall conductivity is valid at zero frequency for finite relaxation rate, one can expect that the only way to invalidate the indefinite growth of the response is by the breakdown of the perturbation theory.
Non-perturbative effects come into play when the electric dipole energy $H_{\rm E}=e{\bf r}\cdot {\bf E}$ is comparable to the relaxation energy scale $\hbar\tau^{-1}$~\cite{morimoto2016topological} or the chemical potential $\mu$.
This puts the upper bound on the electric field under which the second-order description is valid.
Let us consider a linearly dispersing Weyl fermion.
From ${\bf r}_{\rm max}\sim k^{-1}_{\mu}\sim \hbar v/\mu$, we obtain the validity regime ${\bf E}<\min(\mu/ev\tau,\mu^2/\hbar v)$ of the perturbation theory.
Therefore, the upper bound of the nonlinear Hall current in the second-order regime, for example, scales as $j=\sigma_{(2)}|{\bf E}|^2\sim (\log \mu) \mu^2$ or $(\log \mu) \mu^4$, which goes to zero as $\mu\rightarrow 0$.
No real logarithmic divergence occurs.
The important point is instead the enhancement pattern of nonlinear responses until reaching the upper bound.

Finally, let us discuss the application of our analysis to the low-frequency behavior of the resonant photovoltaic effects.
Resonant optical responses do not show logarithmically enhanced frequency dependence in type-II semimetals, because optical transitions occur along a closed surface rather than on a hyperbolic surface~\cite{ahn2020low}.
On the other hand, the anisotropic scaling of a double Weyl fermion can help enhance resonant optical responses.
In particular, the $zzz$ component of the bulk photovoltaic conductivity tensor is enhanced by the factor of $\omega^{-1}$ at sufficiently low frequencies in a double Weyl fermion compared to a Weyl fermion.
Therefore, topological semimetals hosting double Weyl fermions near the Fermi level are promising candidates for photodetection and energy harvesting at low optical frequencies, given that their responses do not cancel each other by symmetry.
%We study the large nonlinear resonant optical responses in a double Weyl semimetal in a following paper~\cite{}.


\end{widetext}

\end{document}